\documentclass[acmtog, nonacm, screen, review]{acmart}
\newenvironment{datamaterial}%
{ \vspace{-0.15cm}%
    \small\noindent{\bfseries Availability of Data and Material:}\par%
    \noindent\ignorespaces}%
{ \par\noindent%
\ignorespacesafterend }%
    
\let\oldmaketitle\maketitle
\renewcommand{\maketitle}{%
  \oldmaketitle%
  \thispagestyle{plain}%
  \pagestyle{plain}}

\usepackage{hyperref}
\usepackage[skip=8pt]{caption}
\usepackage{subcaption}
\usepackage{graphicx}
\usepackage{array}
\usepackage{multicol}
\usepackage{multirow}
\usepackage{longtable}
\usepackage{makecell}
\usepackage{enumitem}
\setlist{nolistsep}
\usepackage{color}
\usepackage{xspace}
\usepackage{rotating}
\usepackage{amsmath}
\newcommand{\rulesep}{\unskip\hspace{0.2cm}{\color{red}\vrule}\hspace{0.2cm}\ignorespaces}

\begin{document}
%%
%% The "title" command has an optional parameter,
%% allowing the author to define a "short title" to be used in page headers.

\title[Untangling Attribution and Estimation]{Untangling Carbon-free Energy Attribution and Carbon Intensity Estimation for Carbon-aware Computing}

\author{Diptyaroop Maji}
\affiliation{%
  \institution{University of Massachusetts Amherst}
  \country{USA}
}

\author{Noman Bashir}
\affiliation{%
  \institution{Massachusetts Institute of Technology}
  \country{USA}
}

\author{David Irwin}
\affiliation{%
  \institution{University of Massachusetts Amherst}
  \country{USA}
}

\author{Prashant Shenoy}
\affiliation{%
  \institution{University of Massachusetts Amherst}
  \country{USA}
}

\author{Ramesh K. Sitaraman}
\affiliation{%
  \institution{University of Massachusetts Amherst}
  \country{USA}
}

% %%
% %% The "author" command and its associated commands are used to define
% %% the authors and their affiliations.
% %% Of note is the shared affiliation of the first two authors, and the
% %% "authornote" and "authornotemark" commands
% %% used to denote shared contribution to the research.
% \author{Diptyaroop Maji}
% % \authornote{Both authors contributed equally to this research.}
% \email{dmaji@cs.umass.edu}
% % \orcid{1234-5678-9012}
% % \author{G.K.M. Tobin}
% % \authornotemark[1]
% % \email{webmaster@marysville-ohio.com}
% \affiliation{%
%   \institution{University of Massachusetts Amherst}
%   % \streetaddress{P.O. Box ?}
%   % \city{Amherst}
%   \state{Massachusetts}
%   \country{USA}
%   \postcode{01002}
% }

% \author{Prashant Shenoy}
% \email{shenoy@cs.umass.edu}
% \affiliation{%
%   \institution{University of Massachusetts Amherst}
%   % \streetaddress{P.O. Box ?}
%   % \city{Amherst}
%   \state{Massachusetts}
%   \country{USA}
%   \postcode{01002}
% }

% \author{Ramesh K. Sitaraman}
% \email{ramesh@cs.umass.edu}
% \affiliation{%
%   \institution{University of Massachusetts Amherst}
%   % \streetaddress{P.O. Box ?}
%   % \city{Amherst}
%   \state{Massachusetts}
%   \country{USA}
%   \postcode{01002}
% }

%%
%% By default, the full list of authors will be used in the page
%% headers. Often, this list is too long, and will overlap
%% other information printed in the page headers. This command allows
%% the author to define a more concise list
%% of authors' names for this purpose.
\renewcommand{\shortauthors}{Maji, et al.}

\begin{abstract}
  % Noman's second draft
Many organizations, including governments, utilities, and businesses, have set ambitious targets to reduce carbon emissions for their Environmental, Social, and Governance (ESG) goals. To achieve these targets, these organizations increasingly use power purchase agreements (PPAs) to obtain renewable energy credits, which they use to compensate for the ``brown'' energy consumed from the grid.  However, the details of these PPAs are often private and not shared with important stakeholders, such as grid operators and carbon information services, who monitor and report the grid's carbon emissions.  
This often results in incorrect carbon accounting, where the same renewable energy production could be factored into grid carbon emission reports and separately claimed by organizations that own  PPAs. \textcolor{black}{Such ``double counting''  of renewable energy production could lead organizations with PPAs to understate their carbon emissions and overstate their progress toward sustainability goals, and also provide significant challenges to consumers using common carbon reduction measures to decrease their carbon footprint}. 
% Further, we show that commonly-used carbon reduction measures, such as load shifting, can have the opposite effect of increasing emissions if such measures use inaccurate carbon intensity signals. 
% For instance, users may increase energy consumption because the grid's carbon intensity appears low, even though carbon intensity may be high when renewable energy attributed to PPAs is excluded. 
Unfortunately, there is no consensus on accurately computing the grid's carbon intensity by properly accounting for PPAs. \textcolor{black}{The goal of our work is to shed quantitative and qualitative light on the renewable energy attribution and the incorrect carbon intensity estimation problems.}

\end{abstract}




\maketitle

\begin{datamaterial}
No additional material provided.
\end{datamaterial}

% 2 columns
% Section - introduction
% \vspace{-0.3cm}
\section{Introduction}
\label{sec:introduction}
Many organizations, including governments, utilities, and  businesses, have set ambitious targets for achieving net-zero carbon emissions by 2050 or sooner~\cite{cities-targets, google-carbon-free, utilities-targets, microsoft-carbon-negative, amazon-carbon-neutral, vmware-carbon, facebook-carbon-neutral}. These organizations often rely heavily on the electric grid. \textcolor{black}{The effort to decarbonize the grid has resulted in significant investments in renewable energy and a higher renewable energy penetration in many parts of the world.} However, the energy from the electric grid is unlikely to be completely carbon-free in the near future~\cite{hotair}.  
Consequently, organizations have begun to address their decarbonization goals by ``offsetting'' carbon-intensive grid energy with zero-carbon renewable energy generated at other physical locations and times. 

Many types of carbon offsets exist today~\cite{understand-carbon-offsets}. An example is an annualized offset that involves purchasing zero-carbon energy over a year to match an organization's annual energy consumption. Another common offset type is the purchase of renewable energy certificates from a renewable energy producer to offset the consumption of brown energy. Studies have pointed out that neither offset achieves true decarbonization since both involve local consumption of non-carbon-free energy~\cite{do-carbon-offsets-work, watt2021fantasy}.  

The strictest offsetting approach used today, known as 24/7 carbon-free energy matching, performs location-specific matching on an hourly basis~\cite{epa_24_7_matching}. This offsetting is primarily done through power purchase agreements (PPAs). The renewable energy producer (the seller) and the purchasing organization (the buyer) sign an agreement where the buyer purchases energy at a ``strike'' price and corresponding renewable energy credits. The seller sells the energy in the wholesale market and settles the difference from the strike prices with the buyer. The buying organization separately buys the energy from the electric grid. It is worth noting that PPA agreements are private transactions and, thus, are not visible to the electric grid and other stakeholders. Studies have argued that 24/7 matching encourages the creation of additional renewable capacity in the grid by providing guarantees on the return on investment to developers who install and run these generation facilities~\cite{twenty-four-seven}.

From a consumer's standpoint, whether the consumer is a residential home or a large organization, there is a critical attribution problem of accurately apportioning the finite amount of carbon-free energy in the electric grid across various consumers, which determines their carbon footprint.  As per the greenhouse gas (GHG) protocol, there are two primary ways of attributing carbon-free energy: \emph{location-based} and \emph{market-based}~\cite{scope2-ghg-guidance}. 
\emph{Location-based} attribution assumes that all consumers within an electric grid consume an electricity mix based on all electricity-generating sources. That is, location-based attribution assumes that electrons from various generation sources in the electric grid cannot be distinguished from one another. Hence, carbon-free energy is attributed to the electric grid, and all the consumers use carbon-free energy in proportion to the grid's carbon-free and fossil-based generation. Alternatively, \emph{Market-based} attribution enables consumers to purchase carbon-free electricity via Renewable Energy Certificates (RECs) or PPA contracts and exclusively claim green electricity for their use. Although it is physically impossible to dispatch a particular source to a specific consumer unless there is a direct transmission line from the generation plant to the consumer, the market-based method still allows consumers to choose the source that supplies their electricity.

% assumes a consumer or organization can choose the generation source that supplies their electricity, even though it is technically impossible to route specific electricity generation to specific end-consumers, even if both are on the same grid.  Hence, the attribution is done via market-based accounting mechanisms, where an organization can claim renewable energy exclusively for their needs by purchasing . 

Consequently, the incentives introduced by both approaches are diametrically opposed, which leads to different stakeholders in the electric grid using different approaches for attribution. For example, a company investing in PPAs would prefer \emph{market-based} attribution to count all the renewable energy generated from that source towards its consumption. Conversely, \emph{location-based} attribution better suits smaller organizations or household consumers who do not have enough demand or financial power to purchase PPAs. The lack of consensus on attribution approaches and the hidden nature of PPAs can lead to double counting carbon-free energy use~\cite{holzapfel2023electricity} and incorrect calculation of an organization's carbon footprint.  Specifically, unless done carefully, renewable energy's contribution may be counted once by an organization that has signed a PPA agreement and again as a part of the grid's energy mix lowering the grid's average carbon emissions.   Such double counting and mixing of location- and market-based attribution would likely lead companies to underestimate and understate their carbon emissions. 

These discrepancies can have two key implications. First, it gives an incorrect impression of the decarbonization of the electric grid. The same carbon-free energy is attributed differently through market- and location-based attribution mechanisms, allowing multiple end users to ``claim'' each unit of carbon-free energy. Second, it complicates the design of information services that estimate the carbon intensity of electricity supply and expose it in real-time to end users. Computing researchers have been developing carbon-aware resource management techniques to ``green'' large-scale systems, such as hyperscaler cloud platforms, enterprise data centers, and large GPU clusters for AI model training. This body of research depends on visibility into the grid's carbon intensity variations~\cite{bashir2021enabling}, which have been used to develop scheduling methods, such as spatial and temporal load shifting. The complexity of carbon attribution mechanisms can result in problems with accurate carbon estimation of electricity, which can lead such research and the resulting techniques deployed in production systems to make incorrect carbon optimization decisions. In particular, consumers may increase their demand because the grid's carbon intensity appears low, even though it may be high when excluding renewable energy attributed to specific companies via PPAs. Thus, accurate attribution and estimation are essential for driving carbon-aware computing research and enabling organizations to achieve their decarbonization goals.  

%\vspace{0.3cm}
%Talk about carbon intensity estimation and its impact on carbon-aware load shifting being done today. 

% \noindent
% {\bf Our Contributions.} 
In this paper, we use a data-driven approach to highlight the challenges in carbon-free energy attribution and carbon intensity estimation. In doing so, we make the following contributions. 
\vspace{0.1cm}
\begin{enumerate}[leftmargin=0.7cm, topsep=0cm]
    \item We first discuss various carbon attribution methods from the perspective of consumers, such as residential users and commercial organizations, and discuss how different attribution methods may be suitable for different end users.
    \item We then discuss how these attribution methods directly impact carbon estimation techniques and how lack of visibility into market-driven methods can result in potential double-counting of renewable energy consumption.
    %We use a data-driven approach to highlight the issue of double counting in green energy attribution and its impact on carbon intensity estimation. 
    Our analysis using ElectricityMaps'~\cite{emap} data for 123 regions demonstrates that up to 66.07\% renewable energy can be double-counted, leading to up to 194\% error in carbon intensity estimations. 
    % \item \textcolor{black}{Finally, we conduct a preliminary study analyzing the impact of inaccurate carbon intensity estimates on the carbon savings achieved by carbon-aware electric vehicle (EV) charging. Using carbon intensity data from California, we show that the reported carbon savings are overestimated, and the discrepancy between actual and reported carbon emission reduction can be potentially up to 156\%.}
    \item \textcolor{black}{Finally, we conduct a preliminary study analyzing the impact of inaccurate carbon intensity estimates on the carbon savings by carbon-aware electric vehicle (EV) charging. Using California as a case study, we show that the discrepancy between the reported and actual carbon savings can be up to 156\%.}
\end{enumerate}

\section{Background}
\label{sec:background}
In this section, we provide background on the electric grid and the carbon intensity of electricity, different carbon accounting methods, how companies can claim renewable energy credits using power purchase agreements, and 24/7 matching of carbon-free energy.

\begin{figure}[t]
    \centering
    \includegraphics[width=\linewidth]{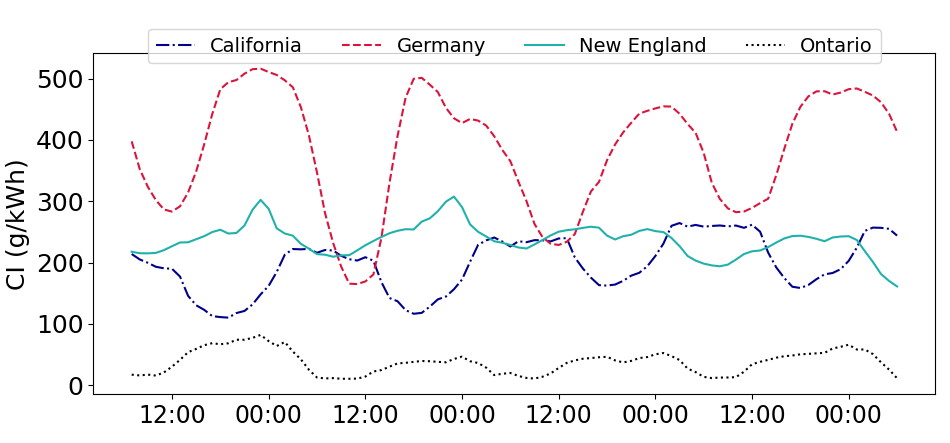}
    \vspace{-0.5cm}
    \caption{\emph{Carbon intensity shows spatial and temporal variations. The regions and periods with high renewable energy have lower carbon intensity.}}
    \vspace{-0.5cm}
    \label{fig:ciso_vs_ercot_ci}
\end{figure}

\subsection{Electricity Distribution Grids}

A region's electricity grid is associated with generating, transmitting, and distributing electricity~\cite{eiaelecdelivery}. Electricity in a region is generated using a mixture of renewable and non-renewable sources. Once generated, the grid transmits electricity over an interconnected network of high-voltage transmission lines. Finally, the electricity is distributed from stations and substations to the end consumers via low-voltage transmission lines. 
\textcolor{black}{Electricity grid operators must ensure that electricity supply always meets demand. The generation of renewable sources like solar and wind is uncontrollable as it depends on the weather. Hence, the grid also maintains a set of generators that can be used to quickly compensate for any demand that renewable sources cannot meet. If the electricity generation in the grid is insufficient to meet the current demand, the grid imports electricity from the grid of neighboring regions with surplus electricity. Thus, the fraction of electricity generated in a region, called the source mix, varies spatially and temporally and often depends upon the neighboring grids' source mix.}

% Since supply should match demand, the grid uses a set of dispatchable generators that can be turned on or off to match a time-varying demand. Sources such as renewable solar and wind
% tend to be intermittent and are assumed to be uncontrolled; other non-renewable sources are then used to meet the remaining demand. Thus, the fraction of electricity generated by each source varies over time and across different regions. 
% Factors like locational marginal price \cite{lmp} and imported electricity also govern the current source mix of a particular region.

\subsection{Carbon Intensity of Electricity}

\textcolor{black}{Carbon intensity is a measure of the greenness of the electricity in a region. The average carbon intensity (CI) of electricity can be defined as the amount of carbon equivalent emitted (in grams) per unit of electrical energy produced or consumed (in kWh). 
% If we consider electricity consumption, then we refer that as the carbon intensity as consumption-based carbon intensity. Electricity consumption is a function of electricity production and imports. Thus, consumption-based carbon intensity depends on the carbon intensity of production and that of imports. 
Calculating the carbon intensity of production, without considering imports, is straightforward and can be mathematically formulated as a weighted average (refer ~\cite{maji2022carboncast}):}

\begin{equation}
    \label{eq:carbonIntensity}
    (Carbon\ Intensity)_{avg}  = \frac{\sum{(E_i * CEF_i)}}{\sum{E_i}}
\end{equation}

where $E_i$ is the electrical energy produced ($MWh$) by a source $i$ \& $CEF_i$ is the carbon emission factor ($g/kWh$) of that source.

\textcolor{black}{The carbon intensity of the consumed electricity in a region is a weighted average of the carbon intensity of the electricity produced in that region and the carbon intensity of the electricity imported from other regions. Calculating the consumption-based carbon intensity is complex since an exporting region may itself import from other regions, and hence we need to trace the flow of electricity back to the original source to get the correct carbon intensity of the imported electricity (refer~\cite{de2019tracking, tranberg2019real, horsch2018flow, tranberg2015power, bialek1996tracing, abdelkader2007transmission, kirschen1999tracing, schafer2019principal, li2013carbon}).}

\textcolor{black}{Figure~\ref{fig:ciso_vs_ercot_ci} shows how the average carbon intensity varies spatially and temporally across four regions. This change in carbon intensity is due to the variability in the mixture of sources generating electricity both across regions and with time. Since non-renewable sources have higher CEFs than renewable sources (refer~\cite{maji2022carboncast}), regions and periods with a higher fraction of electricity generated by renewable sources have lower carbon intensity.}

\vspace{-0.1cm}
\subsection{Scope 2 Carbon Accounting Methods} 

\textcolor{black}{Scope 2 emissions are defined as the indirect greenhouse gas (GHG) emissions that result from the purchase of electricity~\cite{scope2-ghg-guidance}. Scope 2 carbon accounting allows consumers to calculate their operations' scope 2 carbon emissions. Consumers are increasingly trying to reduce the carbon footprint of their electricity consumption as part of their sustainability goals. Few are shifting their demand to low-carbon regions or periods to reduce their emissions; many are setting ``science-based targets (SBTs)''~\cite{bjorn2022renewable, sbt} to align with the global climate goals. Consumers are investing in renewable energy via Power Purchase Agreements (PPAs); SBTs allow them to claim credit for their renewable energy investments and lower the emissions caused by the consumed electricity.} According to the scope 2 GHG guidance protocol~\cite{scope2-ghg-guidance}, there are two methods of estimating carbon intensity and accounting for scope 2 emissions:

\vspace{0.1cm}
\begin{enumerate}[leftmargin=*]
    \item \textbf{Location-based method.} \textcolor{black}{In this method, any electricity generated in a geographical region is attributed to the grid, regardless of any renewable energy investments made by consumers in that grid. All the consumers within that grid have the same carbon intensity, which is estimated using Equation~\ref{eq:carbonIntensity}. The location-based method says that electricity flowing to a consumer is always a mix of all the sources and cannot be segregated based on a particular source.}
    
    % In the location-based method, all the consumers inside a defined geographical location get the same electricity mix in proportion to their consumption. Green energy is attributed to the grid, and the average carbon intensity of the grid mix, including both renewable and non-renewable sources, is used for scope 2 accounting. This is done regardless of any green energy investments made by any specific consumer. Thus, all the consumers in that location share any renewable investment made by a particular company.

    % \vspace{0.1cm}
    \item \textbf{Market-based method.} \textcolor{black}{In this method, consumers investing in renewable energy can claim the ``greenness'' of such electricity while accounting for their carbon emissions, even if they consume electricity from the grid that has a mix of renewable and non-renewable sources. Carbon emissions of electricity consumed by such consumers to meet any remaining demand, or by consumers without investments, are calculated based on the residual grid mix with all electricity under contract removed. The market-based method is a financial way of accounting where electricity can be segregated based on the invested sources, and carbon intensity is different across consumers.}
    
\end{enumerate}

The Scope 2 GHG guide states that all companies should report their emissions using both location-based and market-based methods (``dual reporting''), but they can use either of the methods to meet their carbon emission reduction goals~\cite{scope2-ghg-guidance, brander2018creative}.

% \begin{figure*}[t]
%     \centering
%     \begin{tabular}{cccc}
%     \includegraphics[width=0.23\textwidth]{buildsys23/figures/green-energy-attribution/case1_residential.pdf} &
%     \includegraphics[width=0.23\textwidth]{buildsys23/figures/green-energy-attribution/case2_residential.pdf} &
%     \includegraphics[width=0.23\textwidth]{buildsys23/figures/green-energy-attribution/case3_residential.pdf} &
%     \includegraphics[width=0.23\textwidth]{buildsys23/figures/green-energy-attribution/case4_residential.pdf} \\
%     (a) Neither $H_1$ nor $H_2$ invest in\\ renewable energy through PPA or by buying RECs. &
%     (b) $H_2$ installs rooftop solar,\\ does not sell its RECs, claims them personally. &
%     (c) $H_2$ installs rooftop solar, \\ does not sell its RECs, or uses them personally. &
%     (d) $H_2$ installs rooftop solar, \\ does not sell its RECs, or uses them personally.
%     \end{tabular}
%     \vspace{-0.4cm}
%     \caption{\emph{An illustration of different scenarios that can arise in a residential environment when attributing carbon-free energy to different entities in the electric grid.}}
%     \label{fig:green_energy_attribution}
%     \vspace{-0.6cm}
% \end{figure*}

% \begin{figure}[t]
%     \centering
%     \includegraphics[width=\linewidth]{figures/ppa-figure-v3.pdf}
%     \vspace{-0.7cm}
%     \caption{\emph{An illustration of the power purchase agreement (PPA) commonly used to meet carbon-free energy targets.}}
%     \vspace{-0.5cm}
%     \label{fig:ppa}
% \end{figure}

\subsection{Power Purchase Agreements} 
Power purchase agreements (PPAs) are contracts between an electricity producer (seller) and an electricity consumer (buyer)~\cite{ppa}. PPAs are usually long-term contracts for renewable energy (primarily solar and wind) wherein the buyer can claim renewable energy credits for their investment; that is, they get a certificate and can use that to claim that they have met a fraction of their consumption using the purchased electricity that was generated using carbon-free energy sources. This certificate is called a Renewable Energy Certificate (REC) in the U.S. and Guarantees of Origin (GO) in Europe~\cite{rec_vs_go}. PPAs can be of the following two types:

\begin{figure*}[t]
    \begin{subfigure}{0.31\linewidth}
    \captionsetup{justification=centering}
     \begin{center}
     \includegraphics[width=\textwidth]{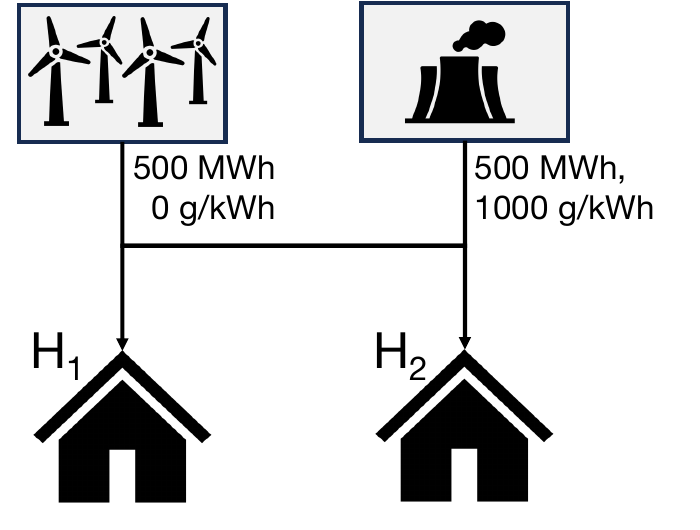}
     \end{center}
     \caption{Neither $H_1$ nor $H_2$ invest in renewable energy through PPA or by buying RECs.}
     \label{fig:gea_case1}
    \end{subfigure}
    \rulesep
    \begin{subfigure}{0.31\linewidth}
    \captionsetup{justification=centering}
     \begin{center}
     \includegraphics[width=\textwidth]{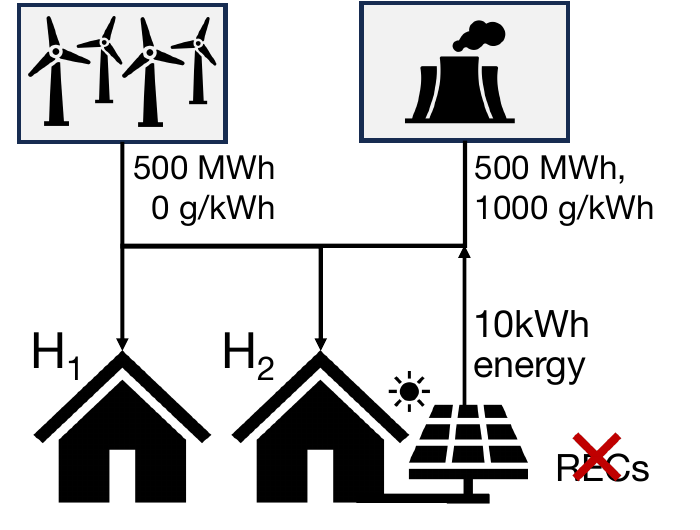}
     \end{center}
     \caption{$H_2$ installs rooftop solar, does not sell its RECs, and uses them personally.}
     \label{fig:gea_case2}
    \end{subfigure}
    \rulesep
    \begin{subfigure}{0.31\linewidth}
    \captionsetup{justification=centering}
     \begin{center}
     \includegraphics[width=\textwidth]{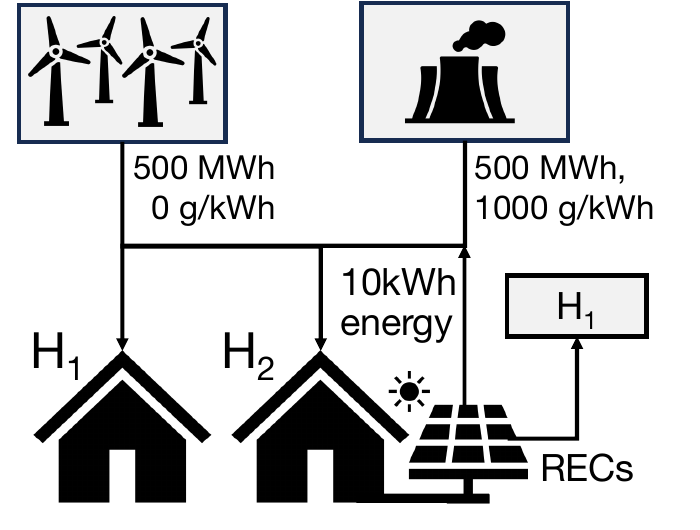}
     \end{center}
     \caption{$H_2$ installs rooftop solar, sells its RECs to an organization that claims them.}
     \label{fig:gea_case3}
    \end{subfigure}
    \vspace{-0.15cm}
     \caption{\emph{An illustration of different scenarios that can arise in a residential environment when attributing carbon-free energy to different entities in the electric grid.}}
     \label{fig:green_energy_attribution}
     \vspace{-0.4cm}
\end{figure*}

\vspace{0.1cm}
\begin{enumerate}[leftmargin=*]
    \item \textbf{Physical PPA:} In a physical PPA, the renewable energy project is either \textit{on-site}, directly supplying the electricity to the buyer, or \textit{off-site}, where the renewable energy is fed into the same grid where the buyer resides~\cite{ppa}. Note that in off-site PPAs, the seller feeds the renewable energy to the grid while the buyer consumes the electricity from the grid. While the seller may agree to ``deliver'' the energy to a specific delivery point in the grid, there is no one-to-one transmission line or another physical method of separating the contracted renewable energy from the rest of the energy consumed by the buyer.

    % \vspace{0.1cm}
    \item \textbf{Financial PPA:} Financial PPAs are financial agreements where the seller can be on the same or a different grid, and electricity is not considered delivered to the buyer. 
    % Figure~\ref{fig:ppa} shows the structure of a financial (or virtual) PPA. 
    In a financial PPA, an organization signs a PPA  with renewable energy producers to purchase green energy for an agreed price. The producer sells the renewable energy in the wholesale electricity market and settles the difference from the agreed price with the buyer~\cite{financial_ppa, financial_ppa_duke}. As a part of this financial transaction, the buyer receives the associated credits and claims renewable energy for its needs. 
\end{enumerate}
% \vspace{0.1cm}
There are other types of PPAs as well, but they typically are a slight variation on either of the two main types of PPAs. One such example is ``sleeved PPA'', which leverages an intermediary utility company to transmit power to a specific delivery point from a renewable energy producer and transfer money from the buyer to the renewable energy producer.

% \begin{figure}[t]
%     \centering
%     \includegraphics[width=\linewidth]{buildsys23/figures/ppa-illustration/ppa-figure-v2.pdf}
%     \vspace{-0.7cm}
%     \caption{An illustration of the power purchase agreement (PPA) commonly used to meet carbon-free energy targets.}
%     \vspace{-0.5cm}
%     \label{fig:ppa}
% \end{figure}

\subsection{24/7 Carbon-Free Energy (CFE) Matching} 

The 24/7 hourly matching of carbon-free electricity (CFE)~\cite{epa_24_7_matching, 24_7_cfe} requires the buyer to consume carbon-free electricity precisely when the seller generated it. The buyer and seller must also be connected to the same grid. For a company to claim they are 100\% carbon-free, its renewable energy investments must align with its demand every hour of the year. 
Currently, the most prevailing carbon offset approach involves market-based matching of renewable energy production with consumption on an annual basis. However, this approach has a drawback: even if renewable energy is not generated in a particular region at a certain time, the buyer can still claim to be using renewable energy as long as the production and consumption average out over the year. The 24/7 matching approach eliminates this issue and enhances accountability in the market-based accounting method. However, it must be noted that 24/7 matching is not the panacea, as even a company that uses 100\% CFE still relies on the grid infrastructure that needs to use fossil-fuel-based energy sources that contribute to carbon emissions.

% Section - green energy attribution
\section{Carbon-free Energy Attribution}
\label{sec:green-energy-attribution}

This section elaborates on how carbon-free energy is attributed to using location-based and market-based methods, with and without renewable investments, by organizations that allow them to claim renewable energy for their needs. We first leverage simple examples of a toy grid to illustrate the challenges in energy attribution under various scenarios in residential as well as commercial settings. While the specifics of renewable energy investments and carbon credits may vary significantly, our analysis captures the essence of the attribution problem for carbon-free energy and demonstrates the potential for discrepancies in accounting for renewable energy. Prior work has also made similar observations~\cite{bjorn2022renewable}, and we contextualize this knowledge for the researchers working on building or computing decarbonization research.

% We will show a few toy examples of carbon-free energy attribution in residential and commercial settings. Note that there can be a myriad of possible scenarios based on the type of PPAs, whether the matching is annual or 24/7, etc. We abstract out all these details and provide a few simple examples. We show how the finite amount of carbon-free energy should be attributed and how that energy may be double counted if the attribution is incorrect. These examples follow the scenarios shown by Bjorn et al.~\cite{bjorn2022renewable}.

\begin{figure*}[t]
    \begin{subfigure}{0.23\linewidth}
    \captionsetup{justification=centering}
     \begin{center}
     \includegraphics[width=\textwidth]{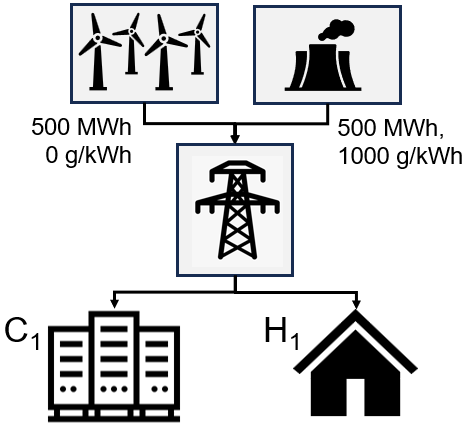}
     \end{center}
     \caption{Neither $C_1$ nor $H_2$ invest in renewable energy through PPA.}
     \label{fig:gea_case1_commercial}
    \end{subfigure}
    \rulesep
    \begin{subfigure}{0.29\linewidth}
    \captionsetup{justification=centering}
     \begin{center}
     \includegraphics[width=\textwidth]{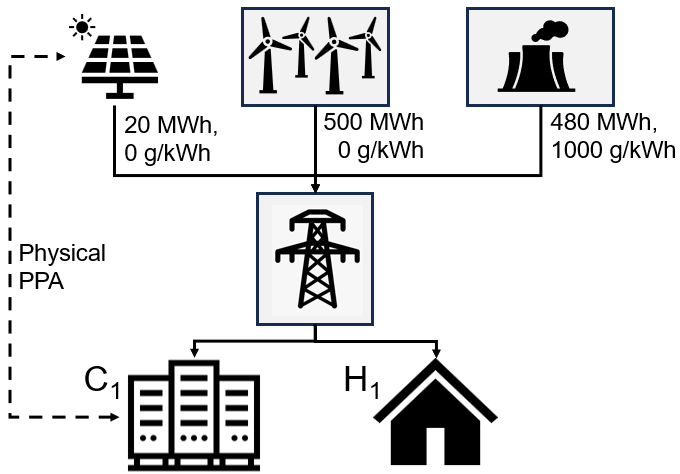}
     \end{center}
     \caption{$C_1$ has physical PPA with offsite solar, and claims renewable energy credits.}
     \label{fig:gea_case2_commercial}
    \end{subfigure}
    \rulesep
    \begin{subfigure}{0.41\linewidth}
    \captionsetup{justification=centering}
     \begin{center}
     \includegraphics[width=\textwidth]{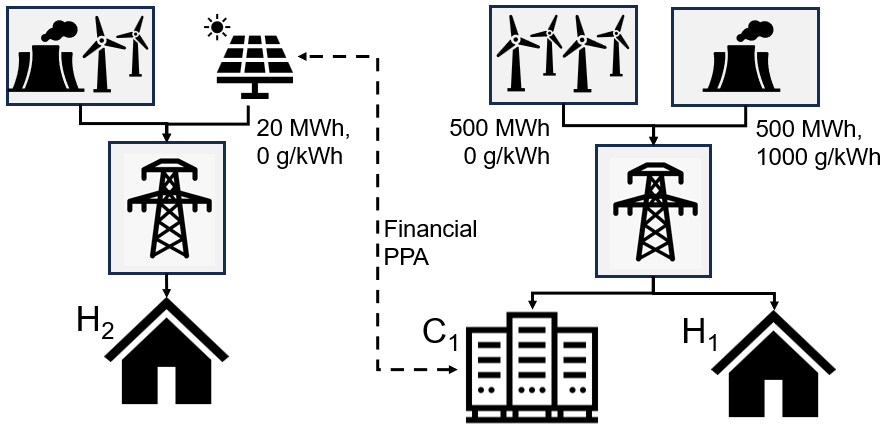}
     \end{center}
     \caption{$C_1$ has financial PPA with solar in a remote grid, and claims renewable energy credits.}
     \label{fig:gea_case3_commercial}
    \end{subfigure}
    \vspace{-0.1cm}
     \caption{\emph{An illustration of different scenarios that can arise in a commercial environment when attributing carbon-free energy to different entities in the electric grid.}}
     \label{fig:green_energy_attribution_commercial}
     \vspace{-0.3cm}
\end{figure*}

\subsection{Residential Settings}
\label{sec:gea_residential}
% First, we consider few cases in the residential settings. Suppose a toy regional grid has two sources --- carbon-free wind, and carbon-intensive coal having CI equal to 1000 g/kWh --- supplying electricity to the consumers. Let the grid demand be 1000 MWh, and the sources contribute equally to the electricity supply. We consider two homes $H_1$ and $H_2$ in that region. Suppose each home consumes 20kWh of electrical energy.
First, we examine a few scenarios for carbon-free energy attribution in residential settings. We have a toy regional grid powered by carbon-free wind energy and carbon-intensive coal power plants. The carbon intensity of the coal power plant is 1000 g/kWh. These sources provide electricity to the consumers with a total grid-level energy demand of 1000 MWh. Both sources contribute equally to the electricity supply. Within this region, we focus on two example homes, labeled $H_1$ and $H_2$. Each of these homes consumes 20 kWh of electric energy over a given time period, such as an hour. We next describe different attribution scenarios, as depicted in Figure~\ref{fig:green_energy_attribution}, a residential version of the commercial example in \cite{bjorn2022renewable}. 

\vspace{-0.1cm}
\subsubsection{\textbf{Case 1:}} The first case represents the state of affairs for most electric grid homes that are passive electricity consumers. Both homes consume electricity directly from the grid, and neither has bought any RECs or has a PPA agreement, as shown in Figure~\ref{fig:gea_case1}. In this case, location- and market-based methods attribute wind and coal proportionally to both homes. Since wind produces $50 \%$ of the electricity in the region, the consumption mix of both homes has $50 \%$ wind. In this case, there is no cause for the discrepancy or scope of double counting, as the carbon-free energy can only be attributed in one way, regardless of the attribution method.

\subsubsection{\textbf{Case 2:}} In the next case, the home $H_2$ installs an \textit{in front of the meter} 10 kWh solar array, which feeds the electricity it generates into the grid. The grid's energy mix now consists of wind, coal, and solar power plants. While the new solar array can meet half of $H_2$'s demand, its contributions toward greening the grid are small as its clean energy production is negligible compared to the overall grid production. In this case, the amount of carbon-free energy attributed to each home varies based on the attribution method used. Under the market-based method (refer to Figure~\ref{fig:gea_case2}), $H_2$ can claim  10 kWh of solar energy towards its energy demand, even if the solar array is not physically delivering any electricity to $H_2$. The remaining demand is then met equally by wind and coal from the grid. Thus, $H_2$ claims 15 kWh of carbon-free energy --- 10 kWh from on-site solar and 5 kWh from wind energy in the electric grid. On the other hand, even if the grid mix that $H_1$ consumes contains a fraction of the solar energy, it cannot be attributed to $H_1$. Thus, $H_1$ still receives the same amount of wind and coal as in Case 1.
% \textcolor{blue}{[DM] Do we cover behind the meter also, which is invisible to the grid?} 

However, if the location-based approach is followed (refer to Figure~\ref{fig:gea_case2}), all of the carbon-free energy is attributed to the grid. Hence, the grid now produces a combined $500.01$ MWh of carbon-free energy, including wind and solar, shared by both homes. Since solar production is negligible with respect to grid production, each home now consumes $10+\varepsilon$ kWh of carbon-free energy and $10-\varepsilon$ kWh of coal ($\varepsilon << 1$).
As expected, the market-based attribution benefits the home with the installed solar, whereas the location-based investment benefits both the homes, although insignificantly in this case. However, if $H_1$ follows the location-based method but $H_2$ follows the market-based method, there will be a discrepancy in counting as the 10 kWh of energy from solar would be counted towards $H_2$, as well as $H_1$ through grid's energy mix. 

% the 10 kWH of solar will be counted doubly, and this situation must be avoided by following only one attribution method.
\vspace{-0.1cm}
\subsubsection{\textbf{Case 3:}} The third case represents the current scenario of the electric grid in developed countries where residential homes with solar installations can sell their RECs to interested organizations through a broker. For the sake of simplicity, we assume that $H_1$ buys RECs from $H_2$ (refer to Figure~\ref{fig:gea_case3}). The location-based attribution method does not change and considers the grid mix containing coal, wind, and solar, as in Case 2. However, in the market-based method, the additional 10 kWh of carbon-free energy is now attributed to $H_1$ instead of $H_2$. $H_2$ cannot claim this carbon-free energy even though it had originally invested in it, or there will be double counting.

\subsection{Commercial Settings}
\label{sec:gea_commercial}
Next, we consider the case of technological companies having data centers in a region that co-resides with other data centers and residential homes. The following section aims to show how to attribute carbon-free energy when these companies invest in physical or financial PPAs. Others \cite{bjorn2022renewable} have also made similar observations.  We also show that lack of visibility into PPAs or RECs can cause  mis-counting of the invested carbon-free energy in the electric grid. 

\subsubsection{\textbf{Case 1:}} Consider the same grid as earlier. The electricity consumers are a company $C_1$ having a 20 MWh data center, and $H_1$ (refer to Figure~\ref{fig:gea_case1_commercial}). Without renewable investment, location- and market-based methods will attribute carbon-free energy proportional to consumption. That is, $C_1$ and $H_1$ will always receive $50 \%$ carbon-free energy, since wind contributes to $50 \%$ of the production. Since there is only one way the energy can be attributed, there is no scope for double counting or discrepancy.

\subsubsection{\textbf{Case 2:}} Suppose $C_1$ invests in an additional 20 MWh \emph{offsite} solar farm that feeds the same grid via a physical PPA (refer to Figure~\ref{fig:gea_case2_commercial}). Assuming the demand is still 1000 MWh, the coal plant reduces its production capacity to 480 MWh. Consequently, the carbon-free production in the grid increases from $50 \%$ to $52 \%$ due to the additional investment. 

In the location-based method, the grid now has $52 \%$ carbon-free electricity, and thus both $C_1$ and $H_1$ have $52 \%$ carbon-free electricity in their consumption mix. However, in the market-based method, $C_1$ can claim the solar energy exclusively and claim to be $100 \%$ renewable since the solar farm meets all its electricity demand. Since 20 MWh is already claimed by $C_1$, the residual grid mix, which represents the mix of energy sources once energy associated with PPAs is removed, now has 500 MWh of wind and 480 MWH of coal. Since $H_1$ did not invest, $H_1$'s electricity demand can only be met by the residual grid mix, which has $51 \%$ carbon-free energy after removing the contracted solar. However, in practice, $H_1$ does not know that $C_1$ has already been claimed when accounting for their consumption mix; they will doubly count the solar energy.

\subsubsection{\textbf{Case 3:}} Finally, suppose $C_1$ instead invests in a financial PPA, where the 20 MWh solar farm is connected to another grid (refer to Figure~\ref{fig:gea_case3_commercial}). This situation is even trickier as the local grid mix does not change as there is no physical energy exchange between the grids, but the remote grid becomes browner when $C_1$ claims the solar energy. In this case, the location-based method would attribute carbon-free energy similar to Case 1. In the market-based approach, $C_1$ can still claim to be $100 \%$ renewable, although there is no physical delivery of electricity from the solar farm to its data center. Contrary to Case 2, $H_1$'s energy mix remains the same, with $50 \%$ of its consumption mix being carbon-free. However, consumers in the remote grid will see their electricity becoming browner and must remove the contracted solar energy when accounting for their fraction of carbon-free energy to avoid any double counting.

\begin{figure}[t]
    \begin{center}
       \includegraphics[width=1\linewidth]{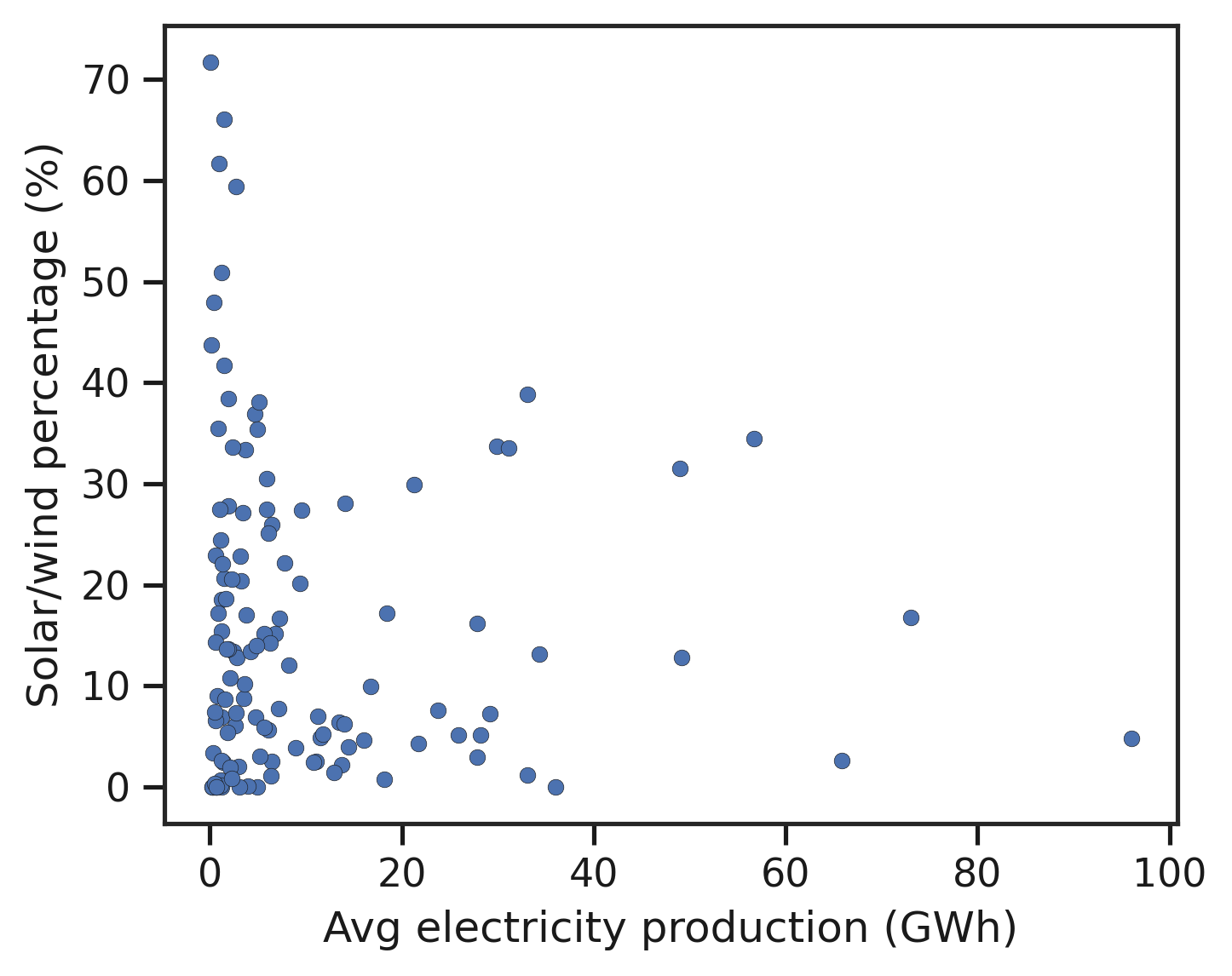}
    \end{center}
    \vspace{-0.4cm}
       \caption{\emph{Hourly average energy production from solar + wind as a percentage of overall energy generation for 123 ElectricityMaps~\cite{emap} regions in 2022. Each circle represents a region.}}
    \label{fig:avg_solar_wind_percentage_emap}
    \vspace{-0.5cm}
\end{figure}

\subsection{Potential for Accounting Discrepancies}
Until now, we have used toy examples and hypothetical situations to explain the different possible scenarios in the electric grid and the potential for double counting and accounting discrepancies in those scenarios. However, the gravity of this situation depends on the prevalence of renewable energy sources, such as solar and wind, in the grid that can potentially be contracted for carbon accounting purposes. We leverage a data-driven approach to assess the current state of the grid and its short-term trend of renewable adoption. We leverage data from electricityMaps~\cite{emap} that provides us with the total energy generation as well as the individual generation from each energy source for 123 regions around the world. For this analysis, we isolate solar and wind energy generation as the two most common renewable energy sources contracted through PPAs. 

Figure~\ref{fig:avg_solar_wind_percentage_emap} shows a scatter plot of total hourly energy production in a given region on $x$-axis and the percentage of generation from solar + wind power plants in the region on $y$-axis. There are two key takeaways from this analysis. First, a significant number of regions have a high renewable energy percentage. If a non-trivial portion of this solar and wind capacity is contracted out through PPAs and if these PPAs are not made visible publicly, there can be a significant potential for accounting discrepancies. In extreme cases, more than 70\% of the total generation mix may be susceptible to this problem. Second, the high energy generation regions use solar and wind for a smaller portion of their energy demand. However, due to their size, even 10\% renewable energy may be much bigger than 60\% for a smaller region in an absolute sense. As a result, all the grid regions, irrespective of their low solar and wind percentage, are susceptible to the attribution problem. 

\begin{figure}[t]
    \begin{center}
       \includegraphics[width=1\linewidth]{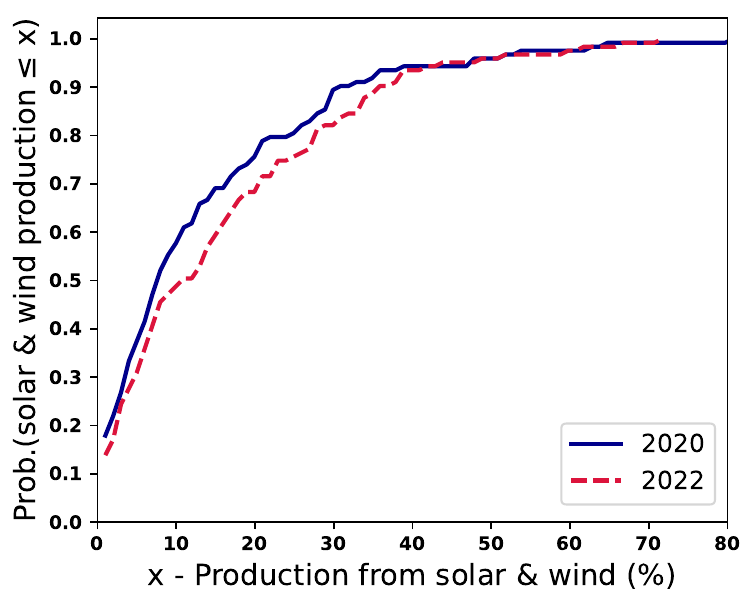}
    \end{center}
    \vspace{-0.25cm}
       \caption{\emph{Hourly average energy production from solar + wind as a percentage of overall energy generation for the same regions as in Fig.~\ref{fig:avg_solar_wind_percentage_emap} in 2020 and 2022.}}
    \label{fig:avg_solar_wind_increase_from_2020_to_2022}
    \vspace{-0.6cm}
\end{figure}

Figure~\ref{fig:avg_solar_wind_increase_from_2020_to_2022} shows the cumulative density function of energy generation from solar + wind power in 2020 and 2022 for the same set of regions as Figure~\ref{fig:avg_solar_wind_percentage_emap}. Despite the short time duration and the impact of a worldwide pandemic, the percentage of energy from solar and wind increased significantly. As the world grapples with climate change's effects, the recent year's curve (red for 2022) will keep moving downwards, meaning more renewable penetration in the grid, and thus complicating the energy attribution problem. 

\vspace{0.1cm}
\noindent
\emph{\textbf{Key takeaway:} The potential for discrepancies in carbon-free energy attribution is real for residential and commercial settings due to various organizations' lack of consensus and information on renewable investments. For example, according to ElectricityMaps' data, solar and wind contributed to $66.07 \%$ electricity generation in South Australia. If all the energy generated from solar or wind is purchased, all that electricity can be potentially double counted. The current state of the grid and its short-term trends suggests that more regions will be susceptible to the attribution problem.}

% Section - CI estimation wrt location- and market-based methods
\section{Carbon Intensity Estimation}
\label{sec:ci-estimation}
In this section, we show how the different carbon-free energy attribution methods discussed in Section~\ref{sec:green-energy-attribution} estimate the carbon intensity of the total grid mix and showcase the discrepancies in the current carbon intensity estimation approaches due to double counting. We assume that electricity production is always sufficient to meet the demand, and there is no electricity exchange across grids for simplicity. We also only consider operational emissions, where all renewables are carbon-free.

\subsection{Carbon Intensity of the Residual Grid Mix}
First, we start with the carbon intensity of the residual grid mix, which is the source mix after removing all contracted carbon-free energy. The residual carbon intensity is only applicable for the market-based method. Suppose the total electricity production in a grid is $E$, and the amount of contracted carbon-free energy is $E_{cf}$. Then, the residual carbon intensity of the grid is
\begin{equation} \label{eq:residual_ci}
    CI_{res} = \frac{\sum{(E_i * CEF_i)}}{E-E_{cf}},
\end{equation}
where $E_i$ is the electrical energy produced ($MWh$) by a source $i$ that is not contracted ($\sum E_i = E-E_{cf}$).

\subsection{Carbon Intensity of the Total Grid Mix}

% That is, 
% \begin{equation} \label{eq:total_ci}
%     CI_{lb} = \frac{\sum{(E_i * CEF_i)}}{E}
% \end{equation}
% where $\sum(E_i) = E$. 
% There is only one value for grid carbon intensity, and it is the same for everyone who consumes electricity from the grid. 

Once the residual carbon intensity is estimated in the market-based method, the carbon intensity of the total grid mix can be calculated in a similar manner as in \cite{bjorn2022renewable}. The market-based carbon intensity ($CI_{mkt}$) is different for different consumers depending on the amount of electricity they purchased, and can be calculated using
\begin{equation} \label{eq:mb_ci}
    CI_{mkt} = \frac{{(D - D_{cf}) * CI_{res}}}{D}
\end{equation}
where $D$ is the total electricity consumed by the consumer (consumer electricity demand), and $D_{cf}$ is the consumer electricity demand met using the invested carbon-free sources. Thus, if someone did not invest in carbon-free electricity, $CI_{mkt} = CI_{res}$, and for consumers with investments, $CI_{mkt} < CI_{res}$.

Alternatively, in the location-based method, any invested carbon-free energy is also a part of the grid mix. The carbon intensity of the grid electricity $CI_{loc}$ always considers the total grid mix and is the same across all consumers. Thus, it can be calculated using Eq.~\ref{eq:carbonIntensity}.

% \vspace{0.1cm}
% Since the $E-E_{cf} \leq E$ and renewables have zero emissions, $CI_{res} \geq CI_{lb}$. We also have $CI_{mb} \leq CI_{res}$ from Equation~\ref{eq:mb_ci}. Thus, the market-based approach always favors the investors of carbon-free energy over non-investors, as their $CI_{mb}$ is lesser.

% \vspace{0.1cm}

% --- data about the residual grid mix often lags by years~\cite{aib_residual, green_e_residual} --- but this is relatively common in the current scenario~\cite{google_247_cfe}.

% \vspace{0.1cm}
% Discrepancies arise when some entities estimate their carbon intensity using these third-party services, whereas other entities that invested in renewables take green energy credit for such investment, leading to inaccurate carbon intensity estimation due to double counting. Even for one entity, if it claims the credit for its PPAs, but relies on these third party-services for carbon intensity estimation when the PPAs are insufficient to meet its demand, the calculated carbon intensity is incorrect as the invested carbon-free energy is counted doubly for the remainder of the energy needs. 

\subsection{Residual Carbon Intensity in Practice}

\begin{figure}[t]
    \begin{center}
       \includegraphics[width=1\linewidth]{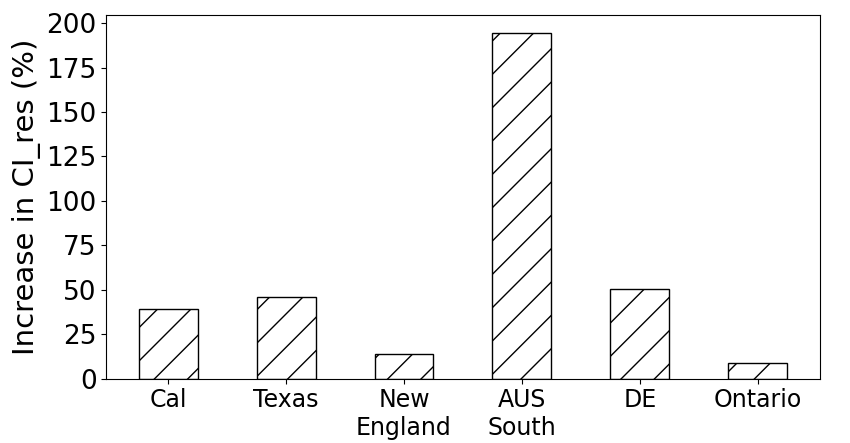}
    \end{center}
    \vspace{-0.3cm}
       \caption{\emph{\textcolor{black}{Carbon intensity of the residual electricity increases as electricity is contracted out. The percentage increase varies across regions, with regions with high solar and wind penetration showing higher increases as all carbon-free energy is contracted out.}}}
    \label{fig:residual_CI_increase}
    \vspace{-0.55cm}
\end{figure}

\textcolor{black}{As explained in Section~\ref{sec:green-energy-attribution}, when a portion of the grid's generation capacity is directly contracted (i.e., purchased) by end consumers, that energy purchase should be directly attributed to the purchasing consumer and not be considered by the grid for computing carbon intensity. Since the residual energy mix is obtained after removing all contracted carbon-free sources, the carbon intensity of the residual grid mix is usually higher than that of the total mix. The magnitude of such increase in a region depends on the amount of renewable penetration as well as the fraction of such renewables that is contracted out. Figure ~\ref{fig:residual_CI_increase} shows the percentage increase in the carbon intensity of the residual mix if all solar and wind energy is contracted, over the total mix if no energy is contracted (carbon intensity data of total grid mix obtained from Electricity Maps~\cite{emap}). Note that when there are no contracts, $CI_{mkt}$ using the total grid mix is the same across all consumers, and is equal to $CI_{loc}$. }

\textcolor{black}{The problem arises when information about the residual mix is unavailable in real-time. Any residential home or commercial organization trying to reduce their carbon emissions relies on third-party services to estimate the carbon intensity of their consumed electricity. However, such services provide $CI_{loc}$, mainly due to the lack of availability of real-time residual grid mix data. On the other hand, consumers who have invested in renewables take green energy credit for such investment and follow $CI_{mkt}$ to account for emissions due to demand met from investments and $CI_{loc}$ to account for emissions due to electricity consumed to meet any remaining demand. This leads to inaccurate carbon intensity estimation for all consumers due to double counting of the invested electricity. These inaccuracies in carbon intensity estimates due to lack of visibility into the residual mix can be quite large due to the differences between $CI_{res}$ and $CI_{loc}$ and can result in significant challenges when performing carbon-aware optimizations.}

% As shown, as the fraction of PPAs in various regional grids grows, so does the carbon intensity of the residual mix. In the case of California's CAISO grid, the carbon intensity becomes higher than that of New England's ISONE grid beyond a certain point. This is due to the high proportion of renewables in the CAISO grid, which makes the residual energy browner faster than the New England grid. 

% The higher carbon intensity of the residual mix also occurs when producers sell renewable energy certificates (RECs). Similar to PPAs, the buyer of these certificates gets the claiming rights to that energy production, and this is the case even when the buyer is not physically connected to the same grid. As an example, even though Electricity Map data \cite{emap_no} shows that the electricity mix in Norway comprises 99\% renewables using location-based method, renewable generators in Norway sell their RECs to consumers in other countries, and the residual mix is only left with 15\% renewable sources~\cite{aib_residual}. Thus, although consumers in Norway are physically supplied with almost carbon-free electricity, they can not claim to have near-zero emissions due to the sale of RECs by their suppliers to prevent double counting. 

\vspace{0.05cm}
\noindent
\emph{\textbf{Key Takeaway:} Lack of information about the real-time residual grid energy mix and standard method of attribution often results in discrepancies while estimating the carbon intensity of the grid. As shown in Figure~\ref{fig:residual_CI_increase}, if all solar and wind electricity in South Australia is purchased, the residual carbon intensity of the grid becomes 370.22 g/kWh, whereas the carbon intensity of the grid, including solar and wind, is only 125.67 g/kWh. This is a 194 \% increase in the grid carbon intensity; if not estimated correctly, these discrepancies can lead to gross overestimation of the ``greenness'' of electric grids. }

\section{Implications for Optimizing Carbon}
\label{sec:ci-implication}
% Sections~\ref{sec:green-energy-attribution} and ~\ref{sec:ci-estimation} explained how current energy attribution and carbon intensity estimation approaches can lead to discrepancies. This section shows the implications of such discrepancies on carbon-aware optimizations performed in data centers, such as spatial and temporal load shifting.  We aim to show how the above-mentioned discrepancies can result in erroneous decision-making or overestimate the actual carbon reductions from such optimizations.

% \subsection{Carbon-aware Optimizations}

\textcolor{black}{Many organizations plan to become net-zero or carbon-negative in the coming decades. The computing industry has been at the forefront of this trend and has announced aggressive net-zero goals. For example, nearly all major public cloud providers have announced plans to run their data centers in a zero-carbon fashion by 2030 or sooner. Cloud providers have also begun exposing their cloud servers' carbon consumption to customers.}

\textcolor{black}{There has been significant research in recent years on how cloud providers and applications can leverage workload flexibility to react to temporal changes in carbon intensity to reduce their carbon footprint further. For example, cloud applications can shift their computations to periods of the day when the carbon intensity is the lowest, an approach known as temporal load shifting~\cite{radovanovic2021carbon}. Others have looked at moving cloud workloads to cloud regions with the greenest electricity, called spatial load shifting~\cite{maji2023bringing}. Load-shifting techniques depend on accurate knowledge of how electricity's carbon intensity varies over time at a given cloud data center and how it differs across the cloud data centers of the cloud provider.}

\textcolor{black}{Our previous sections have highlighted challenges in attributing carbon-free energy generation and estimating grid carbon intensity in the face of location- and market-based mechanisms.  The lack of visibility into market mechanisms can hurt the cloud's carbon optimization efficacy. More generally, as such techniques become more prevalent in varied domains such as green electric vehicle (EV) charging, and flexible load shifting in buildings, the same problems faced by cloud workloads today will be faced more broadly. In this section, we perform a back-of-the-envelope analysis to highlight the impact of these attribution and estimation challenges on a simplified carbon-aware EV charging framework.}

\subsection{Implications on Flexible EV Charging}
\begin{figure}[t]
    \begin{center}
       \includegraphics[width=\linewidth]{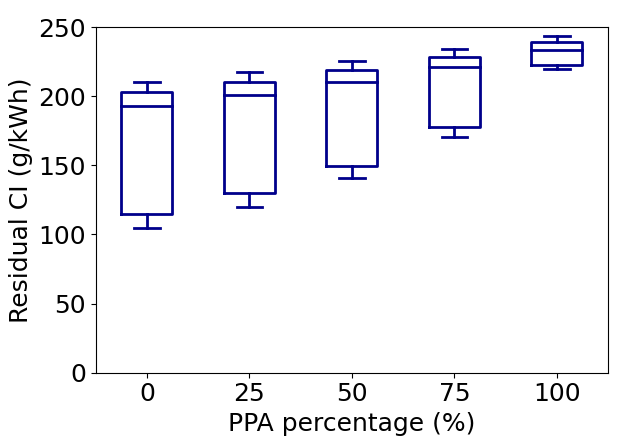}
    \end{center}
    \vspace{-0.3cm}
       \caption{\emph{Temporal variations in carbon intensity decrease as the residual grid electricity gets more brown.}}
    \label{fig:ciso_daily_variation}
    \vspace{-0.6cm}
\end{figure}

% We show how a lack of knowledge about the residual grid mix can also hinder carbon-aware solutions that optimize their demand within a region without involving other regions. 
\textcolor{black}{In the residential sector, people are becoming increasingly carbon-aware and moving flexible loads like electric vehicle (EV) charging, laundry, etc., to low-carbon hours~\cite{huber2021carbon}. However, as the contracted carbon-free electricity sources are removed from the grid, the temporal variations in the carbon intensity decrease progressively. We show this phenomenon for CAISO in Figure~\ref{fig:ciso_daily_variation}. Consequently, there may be a significant discrepancy between the reported and actual carbon savings. For example, suppose Home $H_1$ follows a third-party carbon intensity service to shift doing their laundry from the hour with the highest carbon intensity to the hour with the lowest carbon intensity. In doing so, $H_1$ seemingly reduces their carbon emissions by $39.5 \%$. However, the reduction is only $3 \%$ if all the renewables are already purchased (100\% PPA) and there is no double counting.}% (refer Fig.~\ref{fig:ciso_daily_variation}).

\begin{figure}[t]
    \begin{center}
       \includegraphics[width=\linewidth]{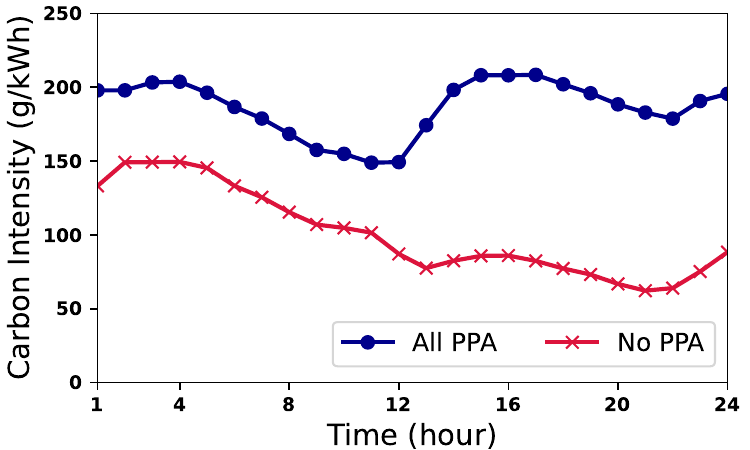}
    \end{center}
    \vspace{-0.2cm}
       \caption{\emph{Carbon-aware load shifting based on a wrong carbon intensity signal may lead to incorrect optimizations and overestimated emissions reduction.}}
    \label{fig:ciso_increase_emissions}
    \vspace{-0.5cm}
\end{figure}

\textcolor{black}{The problem is exacerbated if the third-party service forecasted that the grid carbon intensity will decrease, but the residual grid carbon intensity follows an opposite trend. Figure~\ref{fig:ciso_increase_emissions} shows a 24-hour period in California, where the residual grid carbon intensity increases when the total grid carbon intensity, including solar and wind, is supposed to be less. Suppose Home $H_2$ has a Battery Electric Vehicle (BEV) and a level 2 EV charger~\cite{bev_charging_time}. BEVs take around 10 hours to charge using level 2 chargers~\cite{bev_charging_time}. $H_2$ follows a carbon optimization plan based on third-party forecasts and schedules their EV charging from hour 11 to hour 21. Home $H_1$ then reports an average of $75.7$ grams of carbon emission per kWh consumed electricity. Again, if there were no double counting, and everyone followed the market-based approach, Home $H_2$ would have emitted $194.5$ grams of carbon per kWh of electricity consumed in those hours, $118.8$ grams/kWh more than reported. Not only is this a discrepancy of 156\% carbon emission per kWh, but if $H_2$ had full knowledge about the residual carbon intensity, they may have scheduled their EV charging at different hours.}

\noindent
\emph{\textbf{Key Takeaway:} \textcolor{black}{Incorrect carbon intensity estimates and double counting result in overestimation in the carbon emission reductions reported by organizations following carbon-aware demand response techniques. Although there may still be emission reductions even with incorrect estimates, the situation appears much greener than it is, enabling a false sense of sustainability.}}

\section{Related work}
\label{sec:related_work}
% As the amount of electricity purchased increases, and with the recent push from organizations towards 24/7 matching of carbon-free energy, a few recent works highlight the need for accurate green energy attribution and carbon intensity calculation and show the risks of inaccurate carbon accounting. 

\textcolor{black}{Recently, Bjorn et al.~\cite{bjorn2022renewable} have made similar observations on how inaccurate use of RECs and PPAs provides a sense of inflated emission reduction estimates. We contextualize this knowledge for researchers working on residential or computing decarbonization. We also present a simple case study on how such incorrect estimations affect a decarbonization framework. Brander et al.~\cite{brander2018creative} criticize market-based accounting due to its complexities and lack of accountability and recommend using the location-based method for carbon emission calculations. Our work does not recommend any particular method and just shows the inaccuracies in green energy attribution and carbon intensity estimation and the potential discrepancies between the actual and reported carbon emissions due to that. Holzapfel et al.~\cite{holzapfel2023electricity} show how using location-based and market-based accounting in parallel can lead to double counting in life cycle assessment and offers potential solutions against that. While our work closely relates to that, we highlight such double counting with respect to operational emission accounting residential and commercial computing scenarios.}

\section{Conclusions and Future Work}
\label{sec:conclusion}
% \textcolor{blue}{[DM] This is an old draft. Conclusion needs to be re-written. Will write it once other sections are complete.}

\textcolor{black}{In recent years, organizations have set aggressive goals to reduce the carbon footprint of their electricity consumption as part of their Environmental, Social, and Governance (ESG) goals. To do so, these organizations increasingly use power purchase agreements (PPAs) to obtain renewable energy credits, which they use to compensate for their ``brown'' energy consumption. However, the details of these PPAs are often not shared with grid operators and carbon information services, which monitor and report the grid's real-time carbon emissions. This leads to contracted renewables being double counted, which in turn causes a large discrepancy between the actual and the reported carbon emission savings of an organization --- with the reported savings often overestimating the emission reductions. In this paper, we explain how the current approaches to energy attribution can lead to such discrepancies. We also show with a simplified decarbonization framework how following signals from these carbon intensity services which do not have information about PPAs can lead to an inflated sense of ``greenness''. We hope our work raises awareness in the sustainable computing community about these challenges and leads to better carbon emission accounting. As future work, we plan to do a detailed study of how these discrepancies impact current state-of-the-art carbon optimization techniques and develop algorithms that can work with multiple carbon intensity estimates to reduce carbon footprint without overestimation.}

% Organizations are moving towards 24/7 matching of carbon-free energy and using PPAs as an essential tool to achieve that. In this paper, we highlighted the importance of correctly attributing green energy and calculating CI for carbon accounting in the presence of PPAs. We showed that the CI of the residual mix increases once we account for purchased electricity, and the residual CI of low-carbon regions can sometimes exceed that of even high-carbon regions. Consequently, we showed that carbon-optimization solutions following carbon signals that do not consider PPAs and the residual CI report more carbon savings than actual. We also showed how in some cases, entities can report carbon savings when their carbon emissions would have increased instead under the market-based method.

% We hope that our work raises awareness in the sustainable computing community and leads to the development of newer robust algorithms and more visibility into the grid.

\begin{acks}
% This work is supported in part by NSF grants 2105494, 2021693, and 2020888, and a grant from VMware.
This work is supported in part by NSF grant 2105494 and a grant from VMware.
\end{acks}

% 0.5 columns
\vspace{-0.1cm}
\balance
\bibliographystyle{ACM-Reference-Format}
\bibliography{paper}

% \appendix
% \input{appendix}

\end{document}